# Apparatus and method for plasma processing of SRF cavities

J. Upadhyay,[a] Do Im,[a] J. Peshl,[a] M. Bašović,[b] S. Popović,[a] A.-M. Valente-Feliciano,[c] L. Phillips,[c] and L. Vušković[a]

[a]Department of Physics and Center for Accelerator Science, Old Dominion University, Norfolk, VA 23529, USA

[b]Department of Mechanical Engineering and Center for Accelerator Science, Old Dominion University, Norfolk, VA 23529, USA

[c]Thomas Jefferson National Accelerator Facility, Newport News, VA 23606, USA

**Abstract**

An apparatus and a method are described for plasma etching of the inner surface of superconducting radio frequency (SRF) cavities. Accelerator SRF cavities are formed into a variable-diameter cylindrical structure made of bulk niobium, for resonant generation of the particle accelerating field. The etch rate non-uniformity due to depletion of the radicals has been overcome by the simultaneous movement of the gas flow inlet and the inner electrode. An effective shape of the inner electrode to reduce the plasma asymmetry for the coaxial cylindrical rf plasma reactor is determined and implemented in the cavity processing method. The processing was accomplished by moving axially the inner electrode and the gas flow inlet in a step-wise way to establish segmented plasma columns. The test structure was a pillbox cavity made of steel of similar dimension to the standard SRF cavity. This was adopted to experimentally verify the plasma surface reaction on cylindrical structures with variable diameter using the segmented plasma generation approach**.** The pill box cavity is filled with niobium ring- and disk-type samples and the etch rate of these samples was measured.

## 1. Introduction

The aim of this paper is to describe an apparatus and a method for material removal by plasma assisted reactive ion etching as applied the elliptical SRF cavities, with variable diameter, made of bulk Niobium. The targets of the etching process are also other transient metals and their oxides that are present on the cavity surface as the normal conductive impurities.

The inner surfaces of the superconducting radio frequency (SRF) cavities for particle accelerators are currently chemically treated (etched or electro polished) to remove impurities, reduce the surface resistance of the superconductor, and thus achieve a favorable rf performance. These technologies are based on the use of hydrogen fluoride (HF) in liquid acid baths [1-6], which pose a major environmental and personal safety concern. The plasma etching method would present a much more controllable, less expensive and more environment-friendly processing technology. It is the competitive alternative, which would also provide the unique opportunity to modify the niobium (Nb) surface for energy efficient superconducting rf properties.

The plasma etching method, described in this paper, is based on the cylindrical capacitive rf discharge. Low-frequency rf discharge was chosen because the SRF cavity length is in the order of 1 m and the rf excitation wavelength had to be much longer in order to generate an axially uniform cylindrical rf discharge. In developing the uniform plasma processing technology we have opted for an rf (13.56 MHz) coaxial capacitively coupled plasma of argon (Ar) and chlorine ($Cl_2$) mixture. In this work, the $Cl_2$ was chosen because the niobium and their oxides are converted into volatile chlorides in surface reactions. The presence of negative $Cl_2$ ions makes the plasma electronegative. Moreover, the asymmetry in plasma sheath potential due to surface area difference between the inner and the outer electrode leads to lower ion bombardment energy to the outer electrode wall. The processing of the inner surface of the cylindrical structure requires some form of adjusting the plasma potential to increase the ion bombardment energy on the outer wall and create favorable conditions for the ion assisted etching. In the present methodology, the reversal of asymmetry was achieved by positively dc biasing the inner electrode with the help of an external dc power supply, and by geometrical intervention on the smaller inner electrode. The positive dc bias at the inner electrode, which requires the flow of an additional dc current, not only increases the plasma potential, but also increases the plasma density.

Our approach was to set up the experiment so that the etching effect could be studied for every specific section of the inner surface of the elliptic cavity as a function of each processing parameter. In the first step of process development, we designed a cylindrical cavity with diameter equal to the beam tube of a single cell SRF cavity of 1497 MHz and length 15 cm. Ring shaped Nb samples were placed on the inner surface of this cylindrical cavity for etch rate measurements. The Nb etch rate dependence on the gas pressure, rf power, diameter of the inner electrode, and the concentration of chlorine in $Ar/Cl_2$ mixture was established [7]. The etch rate was improved and the mechanism was improved by varying the temperature of the Nb sample, positive dc bias at the inner electrode and gas conditions [8]. An excessive non-uniformity in the Nb etch rate is observed along the gas flow direction due to depletion of the active radicals [8]. The uniform surface processing across the constant or a varying diameter structure is a complex task even without the depletion of radicals, as observed in [9, 10].

An SRF cavity represents a structure with varying diameter, where the beam tube diameter is much smaller compared to the elliptical section. This imposes some restrictions on the diameter of the inner electrode. In order to create less asymmetric plasma without increasing the diameter of the inner electrode, the shape of the inner electrode was changed and its effect of the self-bias potential was measured. Based on the self-bias potential and etch rate measurement [11, 12] an electrode with a corrugated structure was chosen.

In the third step, we designed a single cell pill box cavity made of stainless steel with similar axial dimensions as the single-cell Nb SRF cavity of 1497 MHz, with multiple mini conflate flanges welded to it for diagnostic purposes. The pillbox cavity was filled with ring type and disk type Nb samples on the inner surface. The purpose was to study the etch rate behavior of Nb on all the available surfaces before etching an actual single cell SRF cavity. The length of the inner electrode was made shorter due to variation of asymmetry in the pillbox cavity. A linear, synchronous translation of the inner electrode and the gas flow inlet was opted to overcome the etch rate non-uniformity [13].

In part 2 we describe the details of the apparatus design and method. The design and images of a pill box cavity with and without ring Nb samples are shown in Section 2.1. The construction of the inner (powered) electrode is described in Section 2.2. The method of segmented plasma generation is outlined in Section 2.3. Electrode and gas inlet manipulation and cell heating techniques are described in sections 2.4 and 2.5. The vacuum rf feedthrough is shown in Section 2.6. The Conclusion contains the most important facts about the apparatus and method.

## 2. Experiment and method

The schematic diagram of the experiment is shown in Fig. 1. This setup has potentially universal application to any concave 3D geometry, but we are describing the method of cavity wall reactive ion etching to remove the transient metal impurities and oxides from the bulk Nb in the surface layer and improve its superconductive properties. The pillbox cavity is representative for the 3D geometrical structure, and it is used to establish plasma etching parameters before the elliptical cell is processed. Work with the pillbox cavity proved useful in defining the process parameters for the reactive ion etching, but the same setup can be applied to any metal surface preparation including thin film deposition and plasma cleaning. Multiple ring samples inside the pillbox cavities were used to determine regions of radical reactions with the surface and to enable their spread over the whole cavity wall. Further, the experiments with various sizes of the inner electrode helped to establish conditions for uniform electrode surface area ratio and determine the sectional step-wise processing procedure. Coaxial rf power feedthrough, filled with atmospheric pressure gas to avoid unnecessary plasma formation, was moved axially together with the inner electrode by a specially designed manipulator.

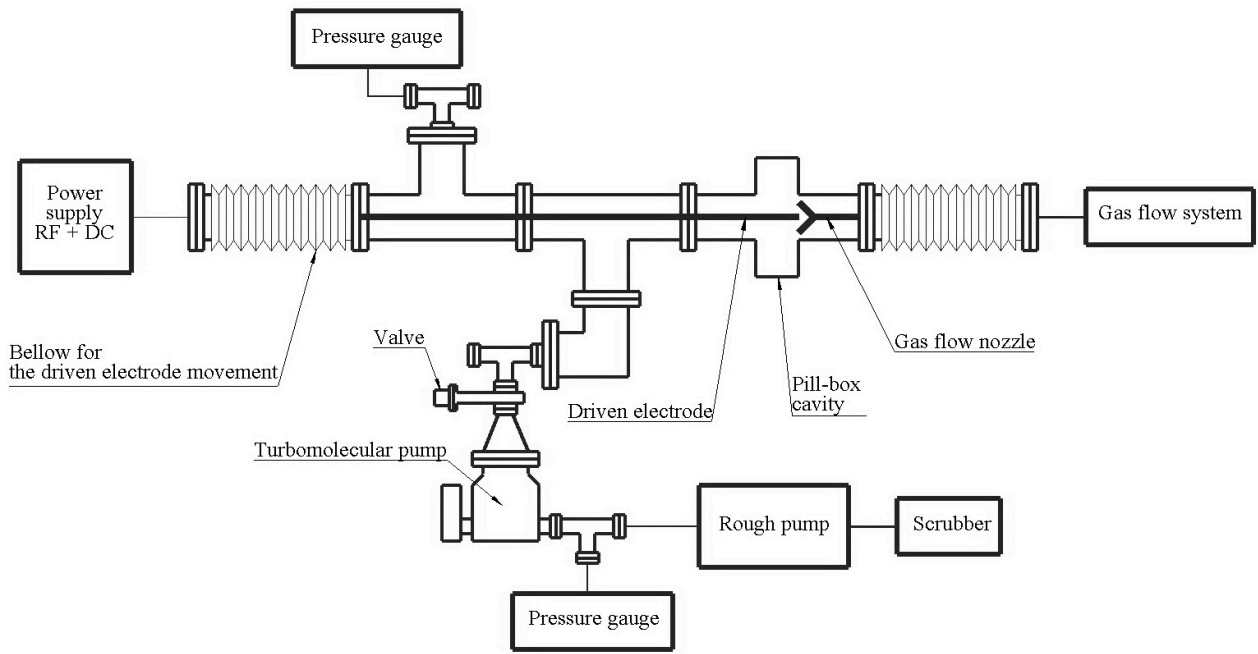

**Fig. 1.** Schematic diagram of the experiment.

The apparatus, shown in Fig. 1, consists of the rf power supply, which is equipped with a matching network and connected in series with a dc power supply that provides positive dc bias on the inner electrode. RF power is coupled to the inner (driven) electrode with a coaxial vacuum feedthrough. The feedthrough and the inner electrode are attached to a controllable axially moving manipulator. The cavity, acting as a vacuum vessel, is connected through the antechamber to the pumping system, which consists of a turbo molecular vacuum pump and a roughing pump with vacuum valves and diagnostic gauges. The exhaust gases are collected and processed in a homemade scrubber that is filled with sodium hydroxide solution in water. Gas is fed to the system through a mixing manifold and a specially designed gas inlet, which disperses the gas mixture. The gas inlet is connected to a second controllable axially moving manipulator, which is synchronized with the first manipulator.

Diagnostic tools are connected through five ports to the pill-box cavity, thus localized observation was achieved. However, in the case of the SRF elliptical cavity, the only observation direction available is through a quartz window at the end of the vacuum assembly and any diagnostic is limited. The cavity wall is electrically grounded and serves as the outer electrode of the cylindrical rf discharge. The surface temperature is controlled by a external heating tape during the etching process.

## 2.1. Single cell pill box cavity

To understand the plasma surface modification on a variable-diameter cylindrical structure we used a single-cell pillbox cavity, shown in Fig. 2. The formerly used accelerator pillbox cavity consisted of an array of cylindrical structures separated by conductive or superconductive walls. In the present case, the pillbox cavity acts as the outer concave grounded electrode. The objective was to study the plasma-surface interaction of all available surfaces on the varied diameter structures.

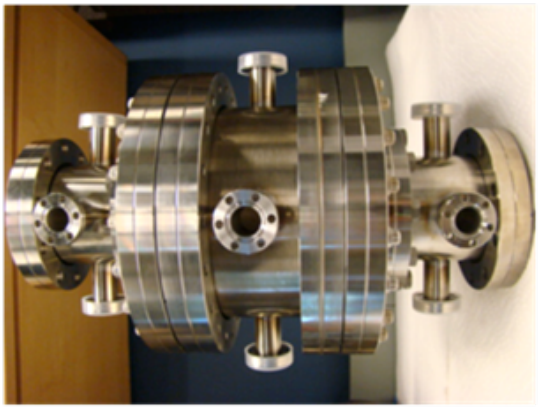

**Fig. 2.** Cylindrical single cell pillbox cavity.

The test pillbox cavity is made of three steel cylindrical structures. The two smaller cylinders have a 7.1 cm inner diameter and 10 cm length. The large cylinder has a 15 cm inner diameter and 10 cm length. The smaller and the larger cylinders mimicked the beam tube and the single cell 1497 MHz SRF cavity. All three cylinders have four holes at 90 degrees each with a mini conflate flange (CF) welded to it for diagnostic and sample holding purposes. These cylindrical structures were joined by transition flanges.

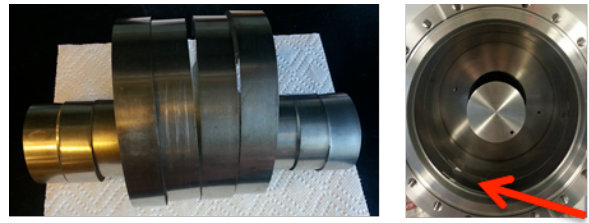

**Fig. 3.** The photos of variable diameter Nb ring samples to be placed in the pill box cavity (left) and Nb ring samples inside the pill box cavity (right).

The inner surface was completely filled with the ring- and disk-type sample to make it as a single cell Nb pill box cavity. In this configuration, each part of the inner structure can be studied separately. The side image of the Nb ring type samples is shown in Fig. 3. The width of each ring sample is 2.5 cm and its ring diameter is equal to the cylindrical steel cavity diameter (15 cm outer diameter and 7.5 cm inner diameter). The disk type samples are used to cover the sidewall of the test cavity to examine the lateral effect of the plasma on Nb.

## 2.2. Electrode geometry

The inner electrode is rf powered and dc biased. Its length is chosen considering complexity in plasma asymmetry due to varied diameter structure of the SRF cavity. In Fig. 4 three options, considered for the inner electrode are shown. The shorter electrode option was chosen to avoid the problem of multiple asymmetries. The plasma sheath potential asymmetry depends on the inner and outer electrodes surface area ratio. In this work, the voltage asymmetry effect was compensated by applying a positive dc bias to the inner electrode. The concern about the applicability of option (a) and (b) was that the separate section exhibit different asymmetry that could not be compensated by a uniform positive bias, which would lead to etching non-uniformity. Therefore, option (c) was chosen and the inner electrode was cut short to 9.0 cm so that it could be confined to a single-diameter cylindrical cavity. In addition, we designed a disc-loaded, corrugated structure to increase the surface area and reduce asymmetry between inner and outer electrode surfaces [12].

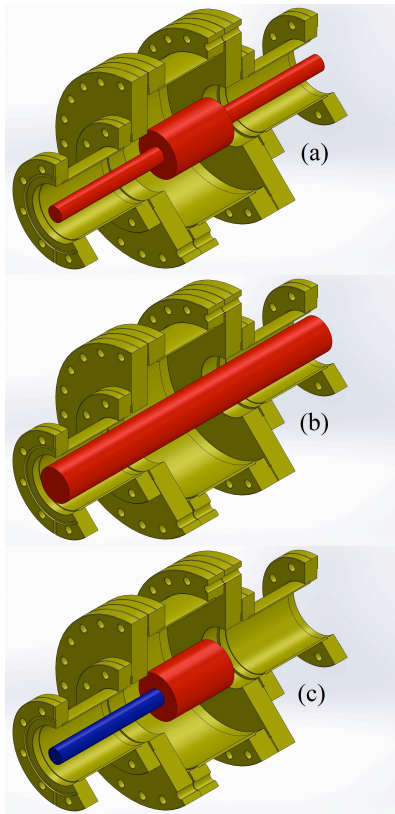

**Fig. 4.** Arrangements of inner electrode length in pillbox cavity (a) variable shape electrode for full length cavity (b) full length electrode (c) shortened electrode in the pill box cavity.

### 2.3. Segmented plasma generation

The Nb ring samples placed nearer to the gas entry point get much more exposure to fresh radicals compared to the sample placed further from the gas entry point. The depletion of active radicals along the gas flow direction due to the consumption creates extreme non-uniformity in etch rate between samples. To overcome this problem, we opted for the segmented plasma production by movement of the inner electrode. The inner electrode is moved, while the plasma is confined to its length in order to establish the plasma surface interaction on a more compact and better controlled surface section.

When the shorter corrugated inner electrode was moved inside the pill box cavity filled with the Nb sample, the sample placed inside the beam tube diameter cavity was etched completely, while the Nb sample placed inside the cell diameter cavity was not etched significantly. The samples placed inside the beam tube diameter cavity are shown in Fig. 5.

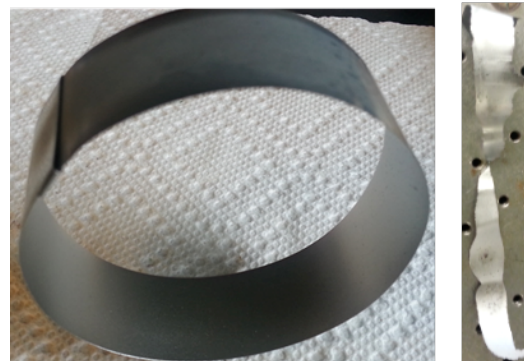

**Fig. 5.** The image of the first Nb sample from the gas flow inlet: before etching (left) and after etching (right).

Therefore, it turned out to be necessary to establish the synchronous step-wise movement of both the electrode and the gas inlet.

### 2.4. Electrode and gas flow manipulation

In the present setup the inner electrode and the gas inlet have been moved so that the gap between them remained constant. Sole sequential movement of the inner electrode could not result in the uniform etching without the gas inlet movment. Even though the inner electrode is further away from the gas flow inlet, due to the low pressure environment the plasma formation around first Nb sample could not be prevented.

The uniform etching of the single cell pillbox cavity is faced with a combination of two challenges. First is the variation in asymmetry, as the varied diameter cell plasma is much more asymmetric than the beam tube diameter plasma. Consequently, the etch rate of Nb may be much higher at the beam tube than at the cell even when the sample was located at a longer distance from the gas flow inlet. The second challenge is in the immediate consumption of the fresh radicals by a Nb sample located nearby.

To overcome these issues, several steps were taken. First, the gas flow inlet shape was changed. It was enclosed in the bellows to have synchronous movement with the inner electrode. The moving gas flow inlet arrangement is shown in Fig. 6.

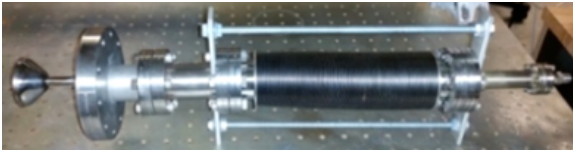

**Fig. 6.** The image (top) and schematic (bottom) of moving gas flow inlet setup.

The gas flow inlet (0.64 cm tube) was placed in a double conical-shape stainless steel funnel to give an outward directional flow to the gas. The diameter of the conical funnel is kept around 5.0 cm (inner electrode diameter) to stop the reflection of gas flow due to obstruction by the inner electrode. The distance between the gas flow inlet and the inner electrode was around 5.0 cm and the inner electrode and the gas flow inlet were moved at the same time.

Step-wise sequential plasma generation enabled the plasma processing of the large diameter cell region only for a longer time, as the beam tube section of the pill box cavity was etched a lot faster. Hence the processing of the cell region consumes most of the etching time, while much less time is needed when the inner electrode was inside the beam tube diameter.

The experiment setup with the inner electrode movement and gas flow inlet movement is shown in Fig. 7.

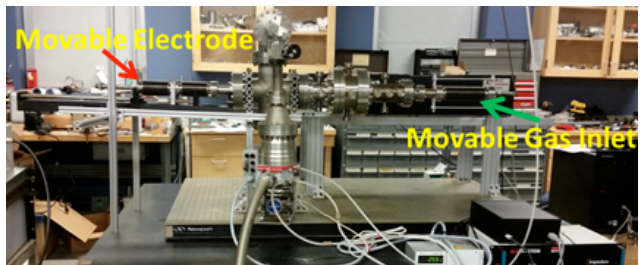

**Fig. 7.** Pillbox cavity with movable inner electrode and movable gas flow inlet.

The simultaneous movement of the gas flow inlet and the inner electrode, the maximum plasma exposure to cell structure, and heating of the cell only helps in reducing the extreme non-uniformity in plasma surface interaction between beam tube diameter and cell structure. With these steps, the etching of the Nb samples placed on the cell diameter cylindrical cavity and the samples placed on the transition flange were possible.

## 2.5. Heating

The etch rate exponentially depends on the temperature [8]. In order to reduce the etch rate on the beam tube and increase the etch rate on the larger diameter cell structure, the heating tape was applied only to the middle section of the cavity with a large diameter. Although the cavity is a metallic structure and the temperature would reach equilibrium, there is a temperature difference between the sections opposite to the inner electrode that interacts with plasma and those not interacting with plasma. The plasma produced in the pillbox cavity is shown in in Fig. 8.

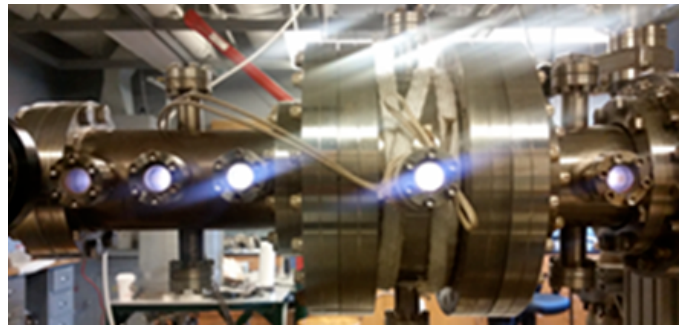

**Fig. 8.** Pillbox cavity with plasma.

Figure 8 indicates that the plasma intensity is higher at the third and fourth window from the left as the inner electrode was positioned at that section as shown in Fig. 9.

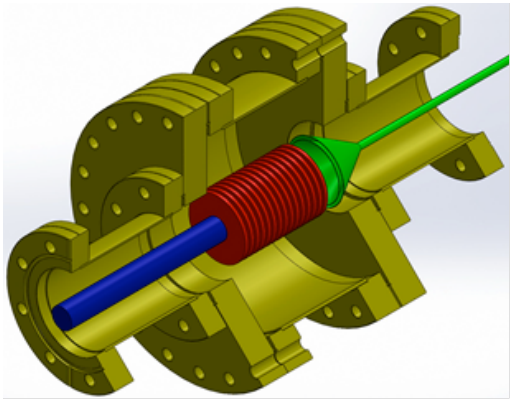

**Fig. 9.** The schematic diagram of the inner electrode, gas flow inlet and pill box cavity during plasma processing.

## 2.6. Atmospheric pressure rf power feedthrough

An atmospheric pressure coaxial rf power feedthrough was developed to couple the rf power and to ensure that the discharge is confined within the cylindrical cavity to be processed. Its schematic is shown in Fig. 10. This rf power feedthrough helps in preventing the discharge inception in front of the cylindrical cavity to be processed. The inner tube of this coaxial feedthrough is 0.64 cm diameter and is made of copper. The outer tube is 1.5 cm in diameter and is made of stainless steel. The ratio of the inner and outer tube is approximately 2.3 to minimize the rf impedance. One end of the feedthrough is connected to an HN type coupler and the other end is connected to an electrical vacuum feedthrough with a mini CF flange. The space between the tubes in the coaxial feedthrough was kept at atmospheric pressure to prevent plasma formation inside the coaxial structure. The vacuum electrical feedthrough is threaded at the end, and is coupled to the powered electrode. The threaded section is covered by the powered electrode to avoid the damage by $Ar/Cl_2$ plasma.

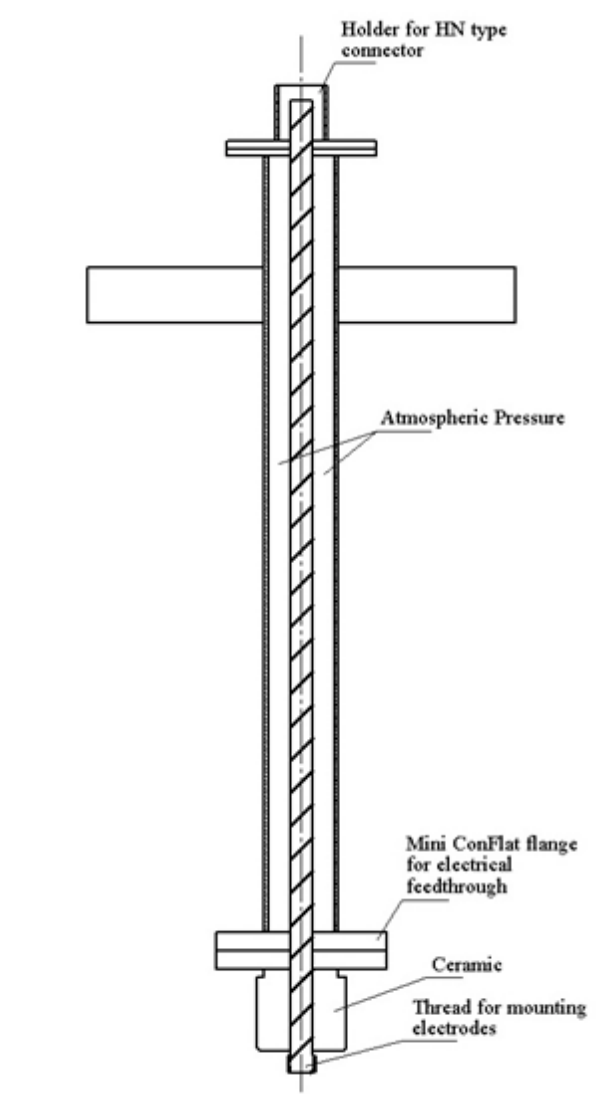

**Fig. 10.** Schematic diagram of atmospheric pressure rf power feedthrough.

## 3. Conclusion

The plasma based surface etching of cylindrical three-dimensional structures is challenging, particularly for the structures with varying shape geometry, including the elliptical SRF cavities. The concept of the pillbox cavity with multiple ports for diagnostic and sample holding purposes has been chosen to test the plasma surface interaction. Capacitive coupled rf plasma of $Ar/Cl_2$ is utilized with the grounded pill-box cavity and powered electrode inside. Here are the most important facts:
(1) The method was developed using a steel pillbox cavity setup that is aligned with

the segmented Nb ring samples. It enables the study of the process parameters and the etching rate for a defined part of the cavity inner surface.
(2) Special emphasis was given to the dimension of the powered electrode to reduce the complexity in plasma asymmetry due to the inner and outer electrodes surface area variation.
(3) Segmented plasma-surface interaction is implemented for efficient and uniform etching process. This was achieved by synchronously moving the powered electrode and the conical shape gas flow inlet. The shape of the gas flow inlet is modified to avoid the reflection from the inner electrode and to act as a barrier for plasma to spread beyond the gas flow inlet.
(4) A non-dissipative, movable atmospheric pressure rf power feedthrough is developed. Its coupling to the powered electrode is established.

The described procedures have proven to be effective for removal of impurity material on elliptical SRF cavities. This same procedure shows promise to become a tool for processing any concave cylindrical surface, including the controlled surface modification and uniform film deposition.

**Acknowledgments**

This work is supported by the Office of High Energy Physics, Office of Science, Department of Energy under Grant No. DE-SC0007879. Thomas Jefferson National Accelerator Facility, Accelerator Division supports J. Upadhyay through a fellowship under JSA/DOE Contract No. DE-AC05-06OR23177.